\documentclass{article}
\def\thru#1{\mathrel{\mathop{#1\!\!\!\!/}}}
\newcommand{\be}{\begin{equation}}
\newcommand{\ee}{\end{equation}}
\newcommand{\Tr}{{\rm Tr}}
\def\bea{\begin{eqnarray}}
\def\eea{\end{eqnarray}}

\begin{document}

\centerline{\LARGE{\bf The Sirens of  Eleven Dimensions }}
\vskip 2cm
\centerline{\Large\bf Pierre Ramond\footnote{\uppercase{T}his research is supported in part by 
the \uppercase{US} \uppercase{D}epartment of \uppercase{E}nergy under grant \uppercase{DE-FG}02-97\uppercase{ER}41029}
}
\vskip .5cm
\centerline{Institute for Fundamental Theory, Physics Department, }  
 \vskip .2cm
\centerline{University of Florida, Gainesville, FL 32611, USA}

\vskip 1.5cm

{\small{\noindent While most theorists are tied to the mast of four dimensions, some have found it irresistible to speculate about eleven dimensions, the domain of  M-theory. We outline a program which starts from the light-cone description of supergravity, and tracks its divergences to suggest the existence of an infinite component theory which in the light-cone relies on the coset $F_4/SO(9)$, long known to be 
linked to the Exceptional Jordan Algebra}}

\section{Souvenirs}
I  met Ian Kogan more than ten years ago. I remember him as being interested in everything,  attacking problems like a cavalry captain, fearlessly and enthusiastically. Growing up in the Soviet Union, he was neither a particle  nor a condensed matter physicist, he was simply put, a physicist. I recall his gracious invitation to lunch in Oxford a few  days after the birth of his child! Ian impressed all around him with his child-like interest in new ideas,  uninhibited  imagination, incredible energy and obvious love of physics.  His premature loss is one of those  events in life that leave everyone who knew him, sadly diminished. 

To honor Ian's memory, I wanted to write something substantive and speculative enough to have aroused his interest.
Thus follows a description of a program that has occupied me for the last seven years. It has not yet reached the stage of obvious physical import,  but  it  has unearthed unique mathematical structures, and raised the specter of a singular theory in eleven dimensions with an infinite number of massless particles of arbitrarily high helicities.

\section{ Eleven Dimensions}
Superstrings seemed to be, up to ten years ago, the most beautiful  theories ever invented since they naturally contain Nature's Interactions, Gravity and Yang-Mills theories, even though they live in ten dimensions and are supersymmetric. From the group-theoretical standpoint, two and ten dimensions are  the natural settings for maximum confusion between fermions and bosons; two dimensions because there is no transverse little group to differentiate between them and ten dimensions because of the triality of $SO(8)$, the transverse little group.

Flat eleven-dimensional space-time is the locale for $N=1$ Supergravity\cite{CJS}, the largest supersymmetric local field theory with maximum helicity two (when reduced to four dimensions),   widely thought to have ultraviolet divergences. It came as a surprise when evidence for much more structure in eleven dimensions was first proposed\cite{WITTEN}: in eleven dimensions lurks a theory that  links superstring theories through various compactifications and dualities, and whose infrared limit is $N=1$ Supergravity. 

Just like strings imply two-forms\cite{KR}, one may have thought  the natural extended object in eleven-dimensions to be a membrane, since $N=1$ supergravity contains a three-form. Unfortunately, membranes seem impossible to quantize, unles there is a (so far unknown) holographic principle to relate membranes to strings.  Glimpses of  M-theory exist in the litterature, and its shadows on lower-dimensional manifolds are  very well defined, but they do not provide enough information to reconstruct the theory in eleven dimensions.  There must be something very special in eleven dimensions that relates bosons and fermions, and evidence for it might be present in its light-cone little group $SO(9)$. 

Perhaps the most interesting feature of $N=1$ supergravity is that it diverges in the ultraviolet. Historically, attempts to understand  divergences have proven vey useful, either as generating operators at small distances in the spirit of effective local field theories, or in pointing to spontaneous breakdown of symmetries in the formulation of the Standard Model. Supergravity divergences  are presumably tamed by M-theory, and we  hope that in learning their structure, we might learn something about M-theory itself.

The one local field theory in four dimensions without ultraviolet divergences\cite{FINITE} is $N=4$ Super-Yang-Mills, which is a compactified form of $N=1$ Super-Yang-Mills in ten dimensions. Curtright conjectured\cite{CURTRIGHT} that the cancellation of its divergences between bosons and fermions was directly related to its ten-dimensional origin and the triality of the transverse little group $SO(8)$.  As $N=8$ Supergravity in four dimensions is the compactified form of $N=1$ supergravity in eleven dimensions, he went on to attribute its divergences to 
group-theoretical properties $SO(9)$ representations of that theory.

\section{Light-Cone Description of $N=8$ Supergravity}
Since we think that evidence for  structure can be found in that of the massless little group, it is convenient to discuss $N=1$ Supergravity in eleven dimensions in terms of light-cone variables. It is sufficient to consider its compactified version to four dimensions where the theory is most economically described with the help of eight Grassmann variables, $\theta^m$, and their conjugates, $\bar\theta^m$, $m=1,2,\dots,8$, each transforming as the real spinor representation of $SO(7)$. The $128$ bosonic fields of $N=1$ supergravity are contained in the  chiral superfield\cite{SWEDES} 

\bea
\Phi(y,\theta)&=&\frac{1}{\partial^{+2}}\,h(y)+\frac{i}{2}\,\theta^m\theta^n\,\frac{1}{\partial^+}\,\bar A^{}_{mn}(y)-\frac{1}{4!}\,\theta^m\theta^n\theta^p\theta^q\,\overline C^{}_{mnpq}\cr
& &\cr
&&+i\,[\,\theta^6]^{}_{tu},\partial^+\,A^{tu}(y)+[\,\theta^8]\,\partial^{+2}_{}\,\bar h(y)\ ,
\eea
where we have used the notation

\be
[\,\theta^6]_{tu}^{}~=~\frac{1}{6!}\,\epsilon^{}_{mnpqrstu}\,\theta^m\theta^n\theta^p\theta^q\theta^r\theta^s\ ,\quad
[\,\theta^8]~=~\frac{1}{8!}\,\epsilon^{}_{mnpqrstu}\,\theta^m\theta^n\theta^p\theta^q\theta^r\theta^s\theta^t\theta^u\ ,
\ee
together with its fermion counterparts, not shown here. In four dimensional language, $h$ and $\bar h$ 
describe the graviton, $A^{mn}$ and $\bar A_{mn}$ the helicity-one and minus one fields , respectively,  
and $\bar C_{mnpq}$ the seventy helicity-zero fields. Complex conjugation  is achieved through Poincar\'e duality

\be
C^{mnpq}_{}~=~\frac{1}{4!}\,\epsilon^{mnpqrstu}_{}\,\bar C^{}_{rstu}\ .
\ee
The fields are only functions of  the light-cone space-time  variables  
\bea
y~=~\,(\,x,\,{\bar x},\,{x^+},\,y^-_{}\equiv {x^-}-\,\frac{i}{\sqrt 2}\,{\theta_{}^m}\,{{\bar \theta}^{}_m}\,)\ ,
\eea
and  form the $\bf 44$ and $\bf 84$-dimensional representations of $SO(9)$, while the fermions label its $\bf 128$-dimensional representation.

The pattern of ultraviolet divergences is related to the group theoretical properties of the light-cone little group, and it has been known for sometime\cite{CURTRIGHT} that the invariants of these three representations satisfy interesting relations.  $SO(9)$ has rank four and thus four Dynkin indices, of order $I^{(n)}_{irrep}$, with $n=2,\,4,\,6,\,8$, ($n=0$ gives the dimension of the representation). Explicitly
\bea
I^{(n)}_{\bf 128}~-~I^{(n)}_{\bf 44}~-~I^{(n)}_{\bf 84}&=&0\ ,\qquad n=0\ ,2\ ,4\ , 6\ ,\cr
& &\cr
I^{(8)}_{\bf 128}~-~I^{(8)}_{\bf 44}~-~I^{(8)}_{\bf 84}&=&-192\ .\eea
The case $n=0$ reflects equal number of  fermions and  bosons, the hallmark of supersymmetry, but the others are more revealing.  Curtright's  conjecture is that the incomplete cancellation of the eight-order indices is the root cause of the divergences of $N=8$ Supergravity. 

Magically, $SO(9)$ contains an infinite set of three irreducible representations\cite{PENGPAN},  called {\it Euler Triplets},  with the similar relations.

\section{Euler Triplets}
Euler Triplets are sets of three representations of $SO(9)$ with the same relations among themselves as the three representations of  Supergravity in eleven dimensions.  While the simplest Euler triplet describes the physical degrees of freedom of Supergravity with spin no higher than two, all others  contain  higher spin fields. In a relativistic theory, they represent massless particles. As a result  their use in a covariant and local physical interacting theory is severely limited\cite{DIFFICULT} by well-known no-go theorems: any finite number of Euler triplets cannot lead to an acceptable interacting physical theory. Fortunately there is an infinite number of them, leaving open the possibility of a non-local theory. In addition,  Euler triplets are naturally expressed in terms of light-cone coordinates, so that  Lorentz invariance may not be manifest until the full set is brought in. 

Euler triplets $\{\,{\bf A}\ ,{\bf B}\ ,{\bf C}\,\}$, where $\bf A$, $\bf B$, $\bf C$ are representations of $SO(9)$, arise from the three equivalent embeddings of $SO(9)$ into $F_4$\cite{GKRS}. To each representation of $F_4$ corresponds one triplet with properties 
\bea
I^{(n)}_{\bf A}~-~I^{(n)}_{\bf B}~-~I^{(n)}_{\bf C}&=&0\ ,\qquad n=0\ ,2\ ,4\ , 6\ ,\cr
& &\cr
I^{(8)}_{\bf A}~-~I^{(8)}_{\bf B}~-~I^{(8)}_{\bf C}&=&-192\,D_{F_4}^{}\ .\eea
Here $D_{F_4}$ is the dimension of the $F_4$ representation associated with the triplet. In general  the largest representation need not be fermionic, but there is an infinite subset for which the largest representation $\bf A$ is a spinor representation, while the other two are bosonic. This subset (generally not supersymmetric) contains therefore as many bosons as fermions and are the true generalization of the supergravity multiplet.  

In mathematical term, each Euler triplet can also be described  by chiral light-cone superfields\cite{BRX}, but the difference with supergravity is that the fields $h,\,A^{mn},\, C^{mnpq}$ and their conjugates are now functions of internal $F_4/SO(9)$ coset variables. 

Chiral fields that describe Euler triplets satisfy Kostant's equation\cite{KOSTANT}, which fixes the functional form of these fields in terms of the internal variables that label the coset. In terms of the Clifford algebra 

\be
\{\, \Gamma_{}^a\,,\, \Gamma_{}^b\,\}~=~2\,\delta_{}^{ab}\ ,~~a,b=1,2,\dots, 16\ ,\ee
generated by $(256\times 256)$ matrices, Kostant's equation is  

\be
\thru {\mathcal K}\,\Psi~=~\sum_{a=1}^{16}\, \Gamma_{}^a\,T^{a}_{}\,\Psi~=~0\ ,\ee
where $T_a$ are the sixteen $F_4$ generators  in the coset $F_4/SO(9)$. In terms of oscillators\cite{FULTON},

\be
T_a~=~-\frac{i}{{2}}\sum_{\nu=1}^3\left\{ (\gamma_i)^{ab}\left(A^{[\nu]\dag}_iB^{[\nu]}_b-B^{[\nu]\dag}_bA^{[\nu]}_i\right)-\sqrt{3}\left(B^{[\nu]\dag}_aA^{[\nu]}_0-A^{[\nu]\dag}_0B^{[\nu]}_a\right)\right\}\ ,
\ee
where the nine $\gamma_i$ are the $(16\times 16)$ Clifford matrices. These operators are written in terms of three copies of $26$ bosonic oscillators $A^{[\nu]}_0,\; A^{[\nu]}_i,\; i=1,\cdots,9,\; B^{[\nu]}_a,\; a=1,\cdots,16$, 
and their hermitian conjugates. 
Under $SO(9)$, the $A^{[\nu]}_i$ transform as ${\bf 9}$, $B^{[\nu]}_a$  as the spinor ${\bf 16}$, 
and $A^{[\nu]}_0$ is a scalar. They satisfy the commutation relations of bosonic  harmonic oscillators

\be
[\,A^{[\nu]}_i\,,\,A^{[\nu']\,\dagger}_j\, ]~=~\delta^{}_{ij}\,\delta_{}^{\nu\,\nu'}\ ,\qquad [\,A^{[\nu]}_0\,,\,A^{[\nu']\,\dagger}_0\, ]~=~\delta_{}^{\nu\,\nu'}\ ,
\ee
as well as for the $SO(9)$  spinor operators 
\be
[\,B^{[\nu]}_a\,,\,B^{[\nu']\,\dagger}_b\, ]~=~\delta^{}_{ab}\,\delta_{}^{\nu\,\nu'}\ ,\ee
in possible conflict with spin-statistics. Remarkably the Euler triplets with equal number of bosons and fermions are precisely those for which the $B$ operators appear quadratically. This links spin-statistics connection with fermion-boson equality, something we have noticed but not understood\cite{ALL}.  Even though the number of fermions and bosons are equal, the Euler triplets by themselves do not support supersymmetry except for the   supergravity triplet.

The  meaning of the internal coordinates associated with these harmonic oscillators is not clear but there is one very intriguing connection with a structure that appeared in the early days of Quantum Mechanics, the Exceptional Jordan Algebra, derived by Jordan, Von Neumann and Wigner in 1935. 

\section{Divergences}
We have alluded to the relation between the litlle group and divergences. To make it specific, consider
 the well-known one-loop beta function 

\be
\beta~=~\frac{11}{3}\,I^{(2)}_{adj}-\frac{2}{3}\,I^{(2)}_f-\frac{1}{3}\,I^{(2)}_H\ ,\ee
where the $I^{(2)}_{adj,\,f,\,H}$ are the quadratic Dynkin indices associated with the adjoint (for the gauge bosons), with spin one-half Weyl fermions, and with complex spin zero fields, respectively. These ``external" group theoretical factors are given by 

\be
\Tr\,(T^A_r\,T^B_r\,)~=~I^{(2)}_r\,\delta^{AB}_{}\ ,\ee 
where $A,B$ run over the gauge group, and $T^A_r$ are the representation matrices in the $r$ representation. It is often not appreciated that the numerical factors  also stem from group theory, this time that of the light-cone  little group.   In four dimensions, where the massless little group is $SO(2)$ with helicity as the sole quantum number,  Hughes\cite{HUGHES} showed that these coefficients are given by 

\be
\frac{1}{3}\,(1-12 h^2_{})\ ,\ee
where $h$ is the helicity of the particle circulating around the loop, whose square can be viewed as the quadratic Dynkin index of the little group. If one applies this formula to  $N=4$ Super-Yang-Mills, the numerical coefficients are easily traced to the quadratic Dynkin index of the $SO(8)$ the little group in ten dimensions, and Curtright traced its vanishing to the triality of $SO(8)$.  At higher loops, it is easy to see that higher order Dynkin indices are put in play.

Now if we look at $N=8$ supergravity in four dimensions, we expect the numerical coefficients in front of its divergences to be expressed in terms of the Dynkin indices  of its mother theory in eleven dimensions, so that the relevant group is $SO(9 )$. A necessary condition for cancellation between bosons and fermions is that the sum of their Dynkin indices cancel (it may not be sufficient since there are three representations and  cancellation is a linear relation which might not work for subdivergences). Since the Dynkin sum rule does not work for the eighth-order index, this hypothesis points to a divergence in the three loop four-graviton amplitude, which is in accord with arguments based on counterterms\cite{DESER}. 

Following Curtright, we assume that the leak in the eight-order indices causes the divergences. Can the addition of Euler triplets alleviate the situation? Unfortunately, the eighth-order deficit leak is of the {\it same sign} as that of  supergravity, so that any one triplet can only add to the problem (unless the statistics of the triplet members is reversed). On the other hand from the no-go theorems, the number of Euler triplets must be infinite, so that the deficit is written as an {\it infinite} sum, like a trace over the dimensions of $F_4$ representations that appear in the hypothesized structure.

For Euler triplets to be relevant to the divergences and to the physics, they must be associated with an algebraic object that contains an infinite number of $F_4$ representations such that their total (weighted?) sum of their representations vanish, like a character formula

\be
\Tr\,{\bf I}~=~\sum\, D^{}_{F_4}~=~0\ .\ee
Possible algebraic structures are Kac-Moody algebras, or perhaps deformed algebras, with each triplet entering with a power of a root of unity in the sum, with the requirement that their quantum trace vanishes. This  slim lead into the possible structure that organizes the Euler triplets is the only one we have at the moment.

The search for a natural generalization of supergravity in eleven dimensions has led us very far in the mathematical world, but not so  
far in the  physics one. Yet in the process we have uncovered several singular mathematical structures, in particular the need for an algebraic structure that contains an infinite number of $F_4$ representations, and also to the charge space of the Exceptional Jordan Algebra. The appearance of these unique mathematical structures that are very specific to eleven dimensions may well be accidental, but experience in string theory shows that mathematical accidents often lead turn into deep  insights. More work needs to be done to determine if this is the case.

\section{Acknowledgments}
I would like to thank Professors M. Shifman and A. Vainshtein for allowing me the  honor of  contributing to this volume.

\end{document}